# Thermodynamics of free and bound magnons in graphene

Andrew T. Pierce[1*], Yonglong Xie[1,2*], Seung Hwan Lee[1], Patrick R. Forrester[1], Di S. Wei[1], Kenji Watanabe[3], Takashi Taniguchi[4], Bertrand I. Halperin[1], Amir Yacoby[1‡]

[1]*Department of Physics, Harvard University, Cambridge, MA 02138, USA*

[2]*Department of Physics, Massachusetts Institute of Technology, Cambridge, MA 02139, USA*

[3]*Research Center for Functional Materials, National Institute for Material Science, 1-1 Namiki, Tsukuba 305-0044, Japan*

[4]*International Center for Materials Nanoarchitectonics, National Institute for Material Science, 1-1 Namiki, Tsukuba 305-0044, Japan*

[*]These authors contributed equally to this work.
[‡]Corresponding author's email: yacoby@g.harvard.edu

**Symmetry-broken electronic phases support neutral collective excitations. For example, monolayer graphene in the quantum Hall regime hosts a nearly ideal ferromagnetic phase at filling factor ν=1 that spontaneously breaks spin rotation symmetry**[1–3]**. This ferromagnet has been shown to support spin-wave excitations known as magnons which can be generated and detected electrically**[4,5]**. While long-distance magnon propagation has been demonstrated via transport measurements, important thermodynamic properties of such magnon populations—including the magnon chemical potential and density—have thus far proven out of reach of experiments. Here, we present local measurements of the electron compressibility under the influence of magnons, which reveal a reduction of the ν=1 gap by up to 20%. Combining these measurements with estimates of the temperature, our analysis reveals that the injected magnons bind to electrons and holes to form skyrmions, and it enables extraction of the free magnon density, magnon chemical potential, and average skyrmion spin. Our methods furnish a**

**novel means of probing the thermodynamic properties of charge-neutral excitations that is applicable to other symmetry-broken electronic phases.**

The interplay between electron-electron interactions and isospin degeneracy in Landau levels can give rise to symmetry-broken states such as quantum Hall ferromagnets (QHFMs)[1,2]. The spin polarized $v$=1 state in monolayer graphene stands out due to its remarkable magnetic properties[3] and represents a unique platform for exploring charge-neutral spin excitations that may be useful for spintronics applications. Recent transport experiments have shown that voltage biases exceeding the Zeeman energy scale $E_Z=g\mu_B B$ provide enough energy for electrons in the $v$=1 edge channel to flip their spin and scatter out of the edge channel. This transition launches a magnonic excitation that may propagate through the bulk and produce a non-local voltage that can be detected microns away[4,5]. However, measurements of the thermodynamic properties of this magnon system—critical for harnessing these novel charge-neutral excitations—remain outside the reach of both transport studies and conventional direct magnetic sensing owing to the dilute magnetization of the system. Here, we perform local electronic compressibility measurements of the $v$=1 QHFM with a scanning single electron transistor (SET) and examine its response to the presence of magnons. We find that pumping magnons into the system results in a marked reduction of the charge gap, typically of about 15-20%. We argue that the gap reduction is a result of magnons binding with electrons or holes to form skyrmions, which, together with estimates of the temperature, allows us to determine the local magnon chemical potential and free magnon density in the system. The method of extracting thermodynamic properties of magnons introduced in our experiments suggests novel routes toward realizing and probing Bose-Einstein condensation in quantum Hall ferromagnets[6], and is more broadly applicable to other flat-band systems with spontaneously symmetry-broken states.

The device and measurement setup are shown in Fig. 1a-b. The hBN-encapsulated graphene device rests on a standard conducting Si/SiO$_2$ substrate, and a narrow top gate (TG) covers part of the device to enable independent tuning of the local filling factor $v_{TG}$. Fig. 1c shows the two-terminal conductance $G_{2T}$ between contacts 2 and 3 as a function of back-gate filling factor $v$ and DC bias $V_{DC}$ at a magnetic field $B$=11 T. Consistent with previous studies[4,5], for $–E_Z < V_{DC} < E_Z$, we see that $G_{2T}$ exhibits a plateau as a result of the quantization of $\sigma_{xy}$. However, the quantized Hall plateau disappears as soon as $|V_{DC}|$ reaches $E_Z$, signaling the onset of magnon generation and absorption processes.

To study the dependence of the $v$=1 gap on the magnon population, we measure the electron chemical potential as a function of filling factor $\mu(v)$ at each value of $V_{DC}$. Fig. 1d shows two representative measurements of the electron chemical potential $\mu(v)$ near $v$=1, with top-gate voltage $V_{TG}$ = 0. Here, $\mu(v)$ jumps sharply as $v$ passes through 1 due to the $v$=1 gap, with the latter taken to be the maximal excursion of $\mu$. The trace at $V_{DC}$ =10 mV (red curve in Fig. 1d) clearly exhibits a reduced gap compared to that at 0 mV (blue curve in Fig. 1d). A detailed $V_{DC}$ dependence of the gap values (summarized in Fig. 1e) exhibits a striking resemblance to that of the transport behavior. Specifically, the gap begins to change principally when $V_{DC}$ exceeds $E_Z$, initially dropping sharply and reaching a ~20% suppression at the highest biases investigated. The gap reduction shown in Fig. 1d-e, observed at many different locations (see Extended Data Fig. 1), demonstrates the remarkable sensitivity of the $v$=1 gap to the presence of magnons.

An important piece of evidence that the $v$=1 gap suppression observed for DC biases $|V_{DC}|>E_Z$ results from magnon generation and absorption is its dependence on $v_{TG}$. As a consequence of the spin order present in the region under the TG, magnons propagate freely across when $v_{TG} = ±1$, but only weakly for $v_{TG} = 0$ and not at all for $v_{TG} = ±2$. Fig. 2a shows the

AC nonlocal voltage $V_{NL}$ measured across contacts 5 and 6, normalized by the AC bias $V_{AC}$ applied between contacts 2 and 3. In addition to the vanishing nonlocal voltage for $|V_{DC}|<E_Z$, we find that for $|V_{DC}|>E_Z$, no significant signal is detected for $|v_{TG}|>2$ nor for $v_{TG}=0$; on the other hand, a strong non local-voltage is seen for $0<|v_{TG}|<2$, in accordance with the expected transport characteristics shown in previous studies[4,5]. Next, we perform gap measurements near contact 5 as a function of $v_{TG}$ and $V_{DC}$, using the same contacts for magnon generation. Fig. 2b shows the reduction in gap at each $V_{DC}$, determined by subtracting from each point the average of the three traces with $|V_{DC}|<E_Z$ at each $v_{TG}$. As in the case of the transport measurement (Fig. 2a), deviations in the $v=1$ gap are only observed for $|V_{DC}|>E_Z$ and $0<|v_{TG}|<2$. Intriguingly, the bias dependence of the gap and the behavior of $V_{NL}$ appear to define three regimes. First, biases $|V_{DC}|<E_Z$ result in no magnon generation and thus leave the gap intact. Second, for biases $E_Z<|V_{DC}|\lesssim 4E_Z$ the gap is suppressed rapidly and the magnitude of the $V_{NL}$ is large. Finally, for larger biases $|V_{DC}|\gtrsim 4E_Z$ the suppression is more gradual and the magnitude of $V_{NL}$ is vanishingly small. Overall, the observed similarities between the $v_{TG}$ dependence of $V_{NL}$ and the gap unambiguously establish that the gap suppression results from magnon propagation into the bulk.

The first step toward understanding the gap suppression is to identify the nature of the charge excitation associated with the $v=1$ gap in the absence of magnons. Theoretical studies[2,7–12] have proposed that the lowest-lying charged excitations at $v=1$ are finite-size skyrmions, consisting of a single charge $\pm e$ "dressed" by one or more extra overturned spins or a valley texture. While valley skyrmions are believed to set the $v=1$ gap under certain idealized conditions[10–12], we find that they are unlikely to play a role in our observations (see Methods). For skyrmions comprised of flipped spins (referred to as "spin skyrmions"), the excitation energy is determined by the competition between $E_Z$ and the exchange energy: larger skyrmions are favored by the exchange interaction, at the expense of $E_Z$ per flipped spin, resulting in an optimal number of flipped spins of order unity. To illustrate this point, we consider a model of spin skyrmions with $s$ flipped spins[13], whose occupation follows a Boltzmann distribution (see Methods). We also include electron-hole asymmetric Wigner crystal (WC)-like terms in the total energy to account for the regions of negative slope in $\mu$ stemming from the effects of correlation (Fig. 1d)[14], along with an overall Gaussian broadening of $\mu(v)$ to account for disorder. Extended Data Fig. 2 shows the best fits to the data using this phenomenological model. The satisfactory agreement between the fit and the data at many different locations validates our model and allows us to determine the Coulomb energy $E_C$, which sets the overall scale for the skyrmion and magnon energies (see Methods), to be around 21.4 meV. Most notably, we find that <$s$>, the mean number of extra spins carried by a charge excitation, is less than 6% of an electron spin in the absence of injected magnons, establishing that the lowest-lying charge excitation consists of bare electrons and holes.

The observed gap suppression can be naturally captured by extending the phenomenological spin skyrmion model to incorporate the presence of magnons[15] (see Methods), where we describe the magnons by an effective Bose-Einstein distribution with chemical potential $\mu_m$ and electron temperature $T$. Since each magnon represents one flipped spin and therefore one unit of $E_Z$, pumping magnons into the system amounts to externally supplying some of the work needed to flip spins. Assuming there is equilibration between the charge excitations and the free magnons, this results in a reduction of the Zeeman free-energy cost by $\mu_m$ per flipped spin for the spin skyrmion, favoring the formation of skyrmions over bare electrons or holes and thus suppressing the overall charge gap. To compare the predictions of this model with our experiments, we compute the $v=1$ gap as a function of $\mu_m$ and $T$ with the parameters obtained by fitting the measured zero-bias $\mu(v)$ curves (see Fig. 3a and Methods). The results of these calculations indicate that considerable enhancements of $\mu_m$ and $T$ are required to achieve the measured gap suppression at large biases (see constant gap contours in Fig. 3a).

In order to use our model to extract the magnon chemical potential $\mu_m$, an independent estimate of the electron temperature $T$ as a function of DC bias is required. Such an estimate is furnished by a measurement of the longitudinal resistance $R_{xx}$ in the presence of magnon pumping. Keeping contact 2 grounded, we apply an AC bias between contacts 1 and 4 and measure the longitudinal AC voltage $V_{xx}$ across contacts 5 and 6. We emphasize that this measurement of $R_{xx}$ is different than the non-local voltage and is not expected to be directly sensitive to contributions from magnon generation and absorption (see Methods). Strikingly, we find that the measured $R_{xx}$ displays a sudden increase for $|V_{DC}|>E_Z$, indicative of its magnon origin. The comparison of the bias-dependent $R_{xx}$ measurement with a measurement of $R_{xx}$ at zero bias as a function of temperature (Fig. 3c) suggests that injecting magnons into the system results in the electron temperature heating up to approximately 3 K. By finding the best-fit temperature for each $V_{DC}$, we extract quantitative values of the electron temperature $T$ (Fig. 3d), which spans three distinct regimes. In the low-bias regime $|V_{DC}|<E_Z$, no magnons are generated and $T$ remains at base temperature. Between $E_Z$ and approximately $4E_Z$, $T$ increases rapidly as a function of bias. Finally, above ~$4E_Z$, $T$ saturates and once again remains constant to the highest biases investigated. We have performed similar estimates using a variety of circuit configurations, both two- and four-terminal (Extended Data Figs. 3, 4 and 5), which point to a similar range of temperatures.

Estimates of $T(V_{DC})$ and the results of our model calculations allow us to relate the measured gap values to $\mu_m$. Specifically, we determine $\mu_m(V_{DC})$ (Fig. 3e) by matching our measured gap values and $T$ to the simulation results (Figs. 3a and d, see Extended Data Fig. 6 for analysis at another location). As in the case of $T$, the measured gap, and $V_{NL}$, we find that $\mu_m(V_{DC})$ exhibits three separate regimes. At low bias $|V_{DC}|<E_Z$, we have $\mu_m=0$ in accordance with the general properties of the Bose-Einstein distribution. At intermediate bias $E_Z<|V_{DC}|\lesssim 4E_Z$, we observe no increase in $\mu_m$, despite the presence of magnon transport signatures in $V_{NL}$ (Extended Data Fig. 7). Thus, the behavior of the measured gap in the intermediate bias regime can be explained as a result of heating due to the injected magnons without invoking the possibility of skyrmion formation. At high bias, $4E_Z\lesssim|V_{DC}|$, where $T \sim 3$ K, we extract values of $\mu_m$ in excess of zero, as expected in the presence of magnon pumping. We emphasize that the gap suppression observed in this regime cannot be explained by heating alone, as this would require the temperature to continue to increase linearly beyond $V_{DC}=\pm 5$ mV and reach as high as 6 K at $V_{DC}=\pm 10$ mV, in direct contradiction to the temperature estimated from our zero-bias $R_{xx}$ measurements (Figs. 3b-c and Extended Data Fig. 3).

Further insight can be gained by examining the density $n_m$ of equilibrated free magnons obtained from our calculations and the mean number of overturned spins per skyrmion $<s>$ as a function of $V_{DC}$. Fig. 3f shows the extracted $n_m(V_{DC})$ in units of magnons per flux quantum $\phi_0$. In the range $E_Z<|V_{DC}|\lesssim 4E_Z$, the finite $V_{NL}$ and gap suppression measurements demonstrate that magnons are at work (Extended Data Fig. 7). However, we find that $\mu_m$ does not increase in this range and $n_m$ therefore remains negligibly small. We speculate that two possible scenarios may explain this apparent contradiction. One hypothesis is that for $E_Z<|V_{DC}|\lesssim 4E_Z$ there is an additional population of magnons, possibly of very long wavelength, which are not in thermal equilibrium with the electrons, and thus are not captured in the computed $n_m$ despite contributing to $G_{2T}$ and $V_{NL}$. A second, more exotic possibility is that in fact only a very small number of magnons is present in this bias regime, which would imply that a highly efficient mechanism of transport is responsible for the changes in $G_{2T}$ and $V_{NL}$. On the other hand, for $4E_Z \lesssim |V_{DC}|$, a finite population of equilibrated free magnons emerges that appears to scale linearly with $V_{DC}$. We note that for a ~100 μm$^2$ sample at 11 T, the highest equilibrium magnon density of ~$3\times10^{-3}$ per $\phi_0$ corresponds to a total number of equilibrated magnons only of order 300. It is possible that a population of non-equilibrated magnons also persists in this regime. In any case, these observations suggest that the absorption rate of magnons at the contacts may be outpaced by the finite population of free magnons, causing $V_{NL}$ to weaken and $G_{2T}$ to level off at high biases (Extended Data Fig. 7). Finally, the corresponding <s> (Fig. 3g) displays a similar trend as $n_m$ and reaches 3 excess overturned spins at the highest biases, consistent with our overall mechanism of gap suppression. The reason for the change in behavior as $V_{DC}$ exceeds $4E_Z$ is not known.

In a low-density electron system, correlation effects induced by the Coulomb repulsion between carriers can result in negative (inverse) electronic compressibility $d\mu/dn$[16,17], which we observe at $\nu=0+\varepsilon$, $\nu=1\pm\varepsilon$ and $2\pm\varepsilon$ (Figs. 4a-b). The associated correlation energy scale, approximated in our model to leading order by the energy of a classical Wigner crystal $E_{WC} \sim \sqrt{\varepsilon}$, governs the magnitude of the negative compressibility. Intriguingly, we find that the negative-compressibility features at $\nu=1\pm\varepsilon$ respond differently to $V_{DC}$ than those at $\nu=0+\varepsilon$ and $2\pm\varepsilon$, with those at $\nu=1\pm\varepsilon$ being greatly diminished at high bias voltages. The pronounced reduction with increasing bias for $E_Z < |V_{DC}| \lesssim 4E_Z$ is presumably due to heating, but the reduction with increasing bias beyond this point, where the electron temperature is found to be constant, signals that the strength of correlations is suppressed by the presence of magnons. A possible explanation is that formation of skyrmions may decrease the magnitude of the correlation energy, because the electric charge of a skyrmion is more spread out than for a bare electron or hole in the lowest Landau level. Comparison of the averaged negative compressibility for $v=0+\varepsilon$, $v=1\pm\varepsilon$ and $2\pm\varepsilon$ (Fig. 4c and d) shows that only near $v=1$ is it sensitive to $V_{DC}$, providing additional evidence for the magnon origin. Further study is required to fully establish the microscopic mechanism of these effects.

Looking ahead, the methods of measuring $\mu_m$ demonstrated here can be used to map out this important quantity over extended spatial regions. As the gradient of $\mu_m$ is the driving force of magnon currents, such studies may provide further new insights into the nature of magnon transport in the system[18]. One can also envision applying our technique to electronic states with exotic magnetic order. In particular, the $v=0$ state in monolayer graphene has been predicted to support spin superfluidity, in which magnons can propagate without dissipation[6,19]. Our experiments also suggest a novel strategy to effectively reduce $E_Z$, or equivalently the spin-$g$-factor, by increasing the magnon chemical potential $\mu_m$, which can be used to drive spin transitions in complex systems like fractional quantum Hall states[11,20–22]. This raises the possibility of dynamical control of quantum phases analogous to recent pump probe experiments[23], but using magnetic excitations instead of THz. Finally, our combined ability of manipulating and probing magnon chemical potential is immediately applicable to intriguing correlated insulating states recently reported in moiré superlattice systems, which are expected to support electrically-addressable neutral excitations similar to the $v=1$ QHFM[24–26].

## Methods

**Sample preparation.** The device consists of monolayer graphene encapsulated by two layers of hexagonal boron nitride on a *p*-doped Si substrate with a 285 nm layer of SiO$_2$, and was fabricated using a dry transfer technique. A gold top gate was defined using electron-beam lithography and thermally evaporated Cr/Au. The final device geometry was defined by electron beam lithography and reactive ion etching. Edge contacts were made by thermally evaporating Cr/Au while rotating the sample using a tilted rotation stage.

**Measurements.** All measurements were carried out in a $^3$He cryostat with a base temperature of approximately 500 mK. Transport measurements were performed using standard lock-in techniques with a 100 µV excitation with frequencies ranging from 17 to 40 Hz. The temperature dependent measurements were recorded by applying current to a resistive heater located at the $^3$He stage. SET tips were fabricated using the procedure described in Ref. 27. The diameter of the SET is approximately 100 nm, and it was held ~300 nm above the encapsulated graphene. Compressibility measurements were performed using DC and AC techniques similar to those described in Ref. 27. The SET serves as a sensitive detector of the change in electrostatic potential $\delta\varphi$, which is related to the chemical potential of the graphene flake by $\delta\mu = -e\delta\varphi$ when the system is in equilibrium. In the AC scheme used to measure $d\mu/dn$, an AC voltage is applied to the Si back gate to weakly modulate the carrier density of the graphene, and the corresponding changes in SET current are converted to chemical potential by normalizing the signal by that from a small AC bias applied directly to the sample. For DC measurements, an analog PID controller is used to maintain the SET current at a fixed value by changing the tip-sample bias. The corresponding change in sample voltage provides a direct measure of $\mu(n)$.

**Spin skyrmion model.** In order to capture the ν=1 gap suppression by the presence of magnons, we formulate a skyrmion model that takes into account the effects of finite temperature and non-zero charge density on the measured chemical potential $\mu(\nu)$. We assume that both the density of overturned spins and the deviation from ν=1 is small. The energy of an elementary charged excitation at ν=1, $E_s^{e,h}$, is given by

$$E_s^{e,h} = \epsilon_s^{e,h} + \left(s + \frac{1}{2}\right) E_Z,$$

where $e$ and $h$ denote electron-like and hole-like excitations, respectively, $s$ is the number of excess flipped spins ($s = 0$ for a bare electron or hole) bound to the charge, $\epsilon_s^{e,h}$ is the energy of the charged excitation in the absence of Zeeman coupling (i.e. due to Coulomb interaction alone), and $E_Z = g\mu_B B$ is the Zeeman energy. In the absence of Landau level mixing, particle-hole symmetry implies

$$\epsilon_s^e = \epsilon_s^h.$$

The skyrmion energies $\epsilon_s^e$ are taken to be $\epsilon_0^e = 0.6266 E_C$, $\epsilon_1^e = 0.5737 E_C$, $\epsilon_2^e = 0.5438 E_C$, $\epsilon_3^e = 0.5248 E_C$, and $\epsilon_{s>3}^e = \left(0.3133 + \frac{\gamma}{(s+x)^{\frac{1}{2}}}\right) E_C$, where the numbers were taken from finite size calculations by Wójs and Quinn[13] and the expression for $s>3$ is an extrapolation which gives the correct limiting value for infinite $s$, while $\gamma = 0.5320$ and $x = 3.327$ are chosen to fit the values for $s = 2$ and 3. The Coulomb energy $E_C$ is defined as $\frac{e^2}{\varepsilon l_B}$, where $\varepsilon$ is the background dielectric constant and $l_b$ is the magnetic length. Here, however, we treat $E_C$ as an adjustable fitting parameter, for which we obtain values corresponding to choices of $\varepsilon$ in the range 10.0 to 12.4.

If we assume that energy, charge, and $S_Z$ are conserved, then maximizing the entropy leads to Boltzmann distributions for both species of spin-$s$ skyrmion

$$n_s^e = e^{-(E_s^e - s\mu_m - \mu)/T}$$

$$n_s^h = e^{-(E_s^h - s\mu_m + \mu)/T},$$

where $n_s^e$ and $n_s^h$ denote, respectively, the densities of electron-like and hole-like skyrmions with $s$ overturned spins in units of inverse flux quanta, respectively, and we have introduced the electron chemical potential $\mu$ and the chemical potential associated with flipped spins—i.e. the

magnon chemical potential—$\mu_m$. The total electron and hole densities $n_e$ and $n_h$ are then given by

$$n_e = \sum_s n_s^e, \quad n_h = \sum_s n_s^h,$$

and the total charge density, or equivalently the filling factor $\nu$, is

$$\nu = 1 + n_e - n_h.$$

These formulae show that, for fixed total density, both the population of charge carriers with total spin $s$ and the electron chemical potential—and thus the gap to charged excitations—depend on the temperature and the magnon chemical potential.

The above formulation determines the filling factor $\nu$ as a function of $\mu$, $\mu_m$, and $T$ under the assumption that the charged excitations do not interact. However, the experimental $\mu_e(n)$ curves exhibit strong negative compressibility near $\nu=1$, indicating that substantial correlation effects are present. At $T=0$, for sufficiently low carrier densities, in the absence of impurities, electrons or holes are expected to form a Wigner crystal, whose energy per carrier has been calculated to be[14]

$$E_{WC}(\nu) = E_C(-0.782133|\nu-1|^{\frac{1}{2}} + 0.2410|\nu-1|^{\frac{3}{2}} + 0.16|\nu-1|^{\frac{5}{2}})$$

This would give a contribution to the chemical potential at $T=0$ of

$$\delta\mu_{WC}(\nu) = \frac{d(|\nu-1|E_{WC}(\nu))}{d\nu} \approx \frac{3}{2}E_{WC}(\nu)\text{sign}(\nu-1)$$

Although a Wigner crystal is expected to melt at a temperature that is much smaller than $E_{WC}$, it is expected that the energy of the correlated liquid is not much different from that of the Wigner crystal. Thus, we assume that the contribution of the correlation energy to the chemical potential has the phenomenological form

$$\delta\mu_{cor}(\nu) = a_{e,h}\frac{3}{2}E_{WC}(\nu)\text{sign}(\nu-1)$$

where $a_{e,h}$ are parameters we fit to experiment, which we allow to be different in order to reflect the observed asymmetries between electrons and holes in our system. We then calculate the densities of skyrmions at finite temperatures using

$$n_s^e = e^{-(E_s^e - s\mu_m - \mu + a_e E_{WC}(\nu))/T}$$

$$n_s^h = e^{-(E_s^h - s\mu_m + \mu + a_h E_{WC}(\nu))/T}.$$

With these distributions, the equation $\nu = 1 + \sum_s (n_s^e - n_s^h)$ again constitutes a relation $\mu(\nu)$ for given values of $E_C, \mu_m, T$ and $a_{e,h}$, which allows us to extract the thermodynamic parameters as follows. At zero DC bias, there is no magnon pumping and we take $\mu_m = 0$ and $T$ to be the cryostat base temperature at 11 T, $T_{base} \approx 0.8$ K. Thus, the zero-bias chemical potential $\mu(\nu, V_{DC} = 0)$ can be fit to obtain estimates of $E_C, a_{e,h}$, and a phenomenological Gaussian density broadening parameter $\Delta\nu$. Extended Data Fig. 2 shows examples of the zero-bias fit results, together with the fit parameters, which are in excellent agreement with the experimental data. These fit parameters are then carried over to compute $\mu(\nu)$ and calculate the $\nu$=1 gap as a function of temperature and magnon chemical potential (Fig. 3a and Extended Data Fig. 6a). An independent estimate of the electron temperature from transport thus allows us to relate the measured gap value under magnon pumping to the corresponding $\mu_m$. Examples of the resulting estimates of $\mu_m$ are shown in Figs. 3e and Extended Data Fig. 6b, where the error bar is estimated by allowing a 3% difference between the measured and calculated gap together with the propagated error from the temperature estimates.

Knowledge of the magnon chemical potential at each value of DC bias $\mu_m(V_{DC})$, along with the estimate of the electron temperature, allows us to estimate the remaining thermodynamic quantities of interest. The average number of excess flipped spins per skyrmion $<s>(V_{DC})$ may be calculated straightforwardly from the distributions via

$$<s> = \sum_s s(n_s^e + n_s^h)/(n_e + n_h).$$

The total density of free magnons $n_m$ is given by

$$n_m = (N_B)^{-1} \sum_{|\mathbf{k}|<k_{max}} n_\mathbf{k}^m,$$

with $n_{\mathbf{k}}^m$ being the density of magnons at wavevector $\mathbf{k}$ given by the Bose-Einstein distribution

$$n_{\mathbf{k}}^m = \frac{1}{e^{(\epsilon_{\mathbf{k}}^m + E_z - \mu_m)/T} - 1}.$$

where $\epsilon_{\mathbf{k}}^m = 1.2533 E_{\text{ex}}(1 - e^{-\frac{k^2 l_B^2}{4}} I_0\left(\frac{k^2 l_B^2}{4}\right))$ and $I_0$ is a modified Bessel function of the first kind, is the energy of a free magnetoexciton, without the Zeeman contribution[28,29]. The cutoff $k_{\max}$ is determined self-consistently by iterating the condition $k_{\max}^2 l_B^2 = \frac{4\epsilon_0^e}{\epsilon_m(k_{\max})}$, where

$$\epsilon_m = \sum_{|\mathbf{k}| < k_{\max}} n_{\mathbf{k}}^m (\epsilon_{\mathbf{k}}^m + E_z)$$

is the energy density of the free magnons, until convergence is achieved within 0.01%. The reason for our cutoff choice is as follows. For $k l_B \gg 1$, a magnetoexciton consists of an electron and a hole separated by a large distance, $d = k l_B^2$, and its energy will be equal to $2\epsilon_0^e$. In thermal equilibrium at low temperatures, the total number of magnons per flux quantum with $k l_B \gg 1$ will be given by $n_m^> \sim \frac{1}{2} k_{\max}^2 l_B^2 e^{(-2\epsilon_0^e)/T}$, and the associated energy will be $\epsilon_m \sim 2 n_m^> \epsilon_0^e$. On the other hand, if $\mu$ is at the center of the energy gap, we should have $n_e = n_h = e^{(-\epsilon_0^e)/T}$. Equating these quantities, we obtain the result for $k_{\max}$ stated above. This relation is consistent with the observation that for a system of linear size $L$, with just a single magnon present, the requirement $d < rL$, with $r$ a constant of order unity, gives $k_{\max}$ of order $L/l_B^2$.

**$R_{xx}$ and $G_{2T}$ thermometry.** To obtain an estimate of the electron temperature $T$ independent of our compressibility measurements, we perform $R_{xx}$ measurements in the presence of magnon generation using the circuit shown in Extended Data Fig. 3. To generate magnons, a DC bias is applied to contact 3, while contact 2 is grounded; in this case, *no AC modulation is applied to the magnon generation contacts*. Strikingly, the measured $R_{xx}$ as a function of DC bias (Extended Data Fig. 3) displays an abrupt change when the applied DC bias exceeds Zeeman energy, reminiscent of the response observed in the magnon transport experiments *with* AC modulation applied to contact 3 (Fig. 1). However, we emphasize that the change in $R_{xx}$ is not caused by magnon absorption events as in the case of the $V_{\text{NL}}$ signal discussed in the main text, because the AC modulation used for monitoring $R_{xx}$ is not applied to the contacts used for magnon generation. Extended Data Fig. 3c and d show $R_{xx}$ measured using the same circuit with no DC bias applied to contact 3 as a function of temperature. Remarkably, we find good agreement between an $R_{xx}$ trace measured at a given DC bias and that at a given temperature (Fig. 3d, where the error bar is estimated by matching the $R_{xx}$ with DC bias to the temperature $R_{xx}$ up to 5% error), suggesting that the change in $R_{xx}$ at a DC bias greater than Zeeman is equivalent to raising the temperature of the system. Comparing these two $R_{xx}$ measurements therefore allows us to determine the temperature of the system when magnons are pumped into the system and determine $\mu_m$ uniquely.

Alternatively, the two-terminal conductance $G_{2T}$ may be used as a proxy for the temperature instead of the four-terminal $R_{xx}$. Extended Data Fig. 4a and b respectively show the bias- and temperature-dependent two-terminal conductance $G_{2T}$ measured with an AC voltage between contacts 1 and 4 using the circuit shown in Extended Data Fig. 3a. As in the case of the $R_{xx}$ measurements, once the system has heated beyond $T\sim 5$ K, the principal signatures of the quantum Hall effect vanish (in this case, the plateau), thus placing an overestimated but crucial upper bound for the temperature of our system. Extended Data Fig. 4c shows $G_{2T}$ at -10 mV DC bias compared with a selection of zero-bias traces taken at various temperatures, which points to a temperature in the range of ~3 K, in good agreement with the results obtained by analyzing $R_{xx}$. We have verified this behavior in numerous circuit configurations, both in two-terminal and four-terminal configurations, which consistently point to the same range of temperatures (Extended Data Fig. 5, except for positive biases in Extended Data Fig. 5h, which is likely due to a bad contact). We regard the $R_{xx}$ measurements as a more reliable indicator of temperature, as $G_{2T}$ is more susceptible to effects stemming from contact resistance. Nevertheless, our observation that both the $R_{xx}$ and $G_{2T}$ thermometry techniques yield approximately the same electron temperature leads us to conclude that reliable estimates can be derived from either technique.

**Role of valley skyrmions.** The $\nu=1$ quantum Hall state in suspended graphene was originally proposed to support valley skyrmions as its lowest-lying charged excitations[10,11,30]. In encapsulated devices, however, the presence of the boron nitride substrate may break sublattice symmetry and therefore disfavor the formation of valley skyrmions. Although we do not find direct evidence for a gap at the charge neutrality point in our device, we observe a robust incompressible state at ν=5/3, with an incompressible peak comparable in magnitude to that occurring at ν=1/3 and 2/3 (Extended Data Fig. 8). The conspicuous absence of this state in previous local compressibility measurements on suspended devices was attributed to low-lying valley-skyrmionic excitations with energy less than that of a Laughlin quasiparticle[11,27]. Thus, the observation of robust incompressible states at ν=5/3 strongly suggests that valley skyrmions are disfavored in our sample. Moreover, previous thermal activation measurements at ν=1 under tilted magnetic fields on encapsulated devices[3] report an effective g-factor greater than 2, indicating that the lowest charged excitations carry extra flipped spins. Finally, within the spin skyrmion model of gap suppression, we do not expect the presence of magnons to alter the energy cost of adding a valley skyrmion. Hence, we conclude that valley skyrmions are unlikely to play a significant role in the observed ν=1 gap reduction.

**Discussion of possible gap at the CNP.** To search for evidence for sublattice symmetry breaking, we performed high resolution local compressibility measurements near the CNP at zero magnetic field, which we compare to a model that considers the sublattice-gapped Dirac form $\mu(n) = \sqrt{\frac{\Delta_0^2}{4} + \frac{\pi v_F^2 n}{\hbar^2}}$, where $\Delta_0$ and $v_F$ are the sublattice gap and the Fermi velocity respectively. Extended Data Fig. 9 shows two fits of the measured inverse compressibility at zero magnetic field to the sublattice-gapped Dirac model, one with disorder broadening and one without. The un-broadened fit favors a scenario in which with sublattice gap is zero. The broadened fit, however, yields a mean squared error (MSE) approximately one-half that of the unbroadened fit, and favors a scenario in which the sublattice gap is approximately 12.3 meV with a broadening of $7\times10^9$ cm$^{-2}$, consistent with that extracted from our fit to the ν=1 gap at high magnetic field. These considerations suggest that sublattice symmetry is likely broken by the BN substrate, disfavoring the formation of valley skyrmions, despite the compressibility signature of the gap being obscured by disorder broadening at zero magnetic field.

**Acknowledgments**

We acknowledge discussion with P. Jarillo-Herrero. This work was primarily supported by the U.S. Department of Energy, Basic Energy Sciences Office, Division of Materials Sciences and Engineering under award DE-SC0001819. Fabrication of samples was supported by the U.S. Department of Energy, Basic Energy Sciences Office, Division of Materials Sciences and Engineering under award DE-SC0019300. A.Y. also acknowledges the Gordon and Betty Moore Foundations EPiQS Initiative through Grant No. GBMF9468, ARO Grant No. W911NF-14-1-0247 and the STC Center for Integrated Quantum Materials, NSF Grant No. DMR-1231319. A.T.P. acknowledges support from the Department of Defense through the National Defense Science and Engineering Graduate Fellowship (NDSEG) Program. Y.X. acknowledges partial support from the Harvard Quantum Initiative in Science and Engineering. A.T.P, Y.X and A.Y acknowledge support from the Harvard Quantum Initiative Seed Fund. P.R.F. acknowledges support from the National Science Foundation Graduate Research Fellowship under Grant No. DGE 1745303. K.W. and T.T. acknowledge support from the Elemental Strategy Initiative conducted by the MEXT, Japan, Grant Number JPMXP0112101001, JSPS KAKENHI Grant Number JP20H00354 and the CREST (JPMJCR15F3), JST. This work was performed, in part, at the Center for Nanoscale Systems (CNS), a member of the National Nanotechnology Infrastructure Network, which is supported by the NSF under award no. ECS-0335765. CNS is part of Harvard University.

**Author contributions**

A. T. P., Y. X., and A. Y. designed the experiment. A. T. P. and Y. X. performed the scanning SET experiment, the temperature-dependent transport measurements and analyzed the data with input fom A. Y., and S. H. L. fabricated the device and performed the dilution refrigerator transport measurements. B. I. H., A. T .P. and Y. X. performed the theoretical analysis and carried out the numerical calculations. K. W. and T. T. provided hBN crystals. All authors participated in discussions and in writing of the manuscript.

**Competing interests**

The authors declare no competing interest.

**Data availability**

The data presented in this paper are available from the corresponding author upon reasonable request.

**Code availability**

The code that supports the findings of this study is available from the corresponding author upon reasonable request.

# Figure 1

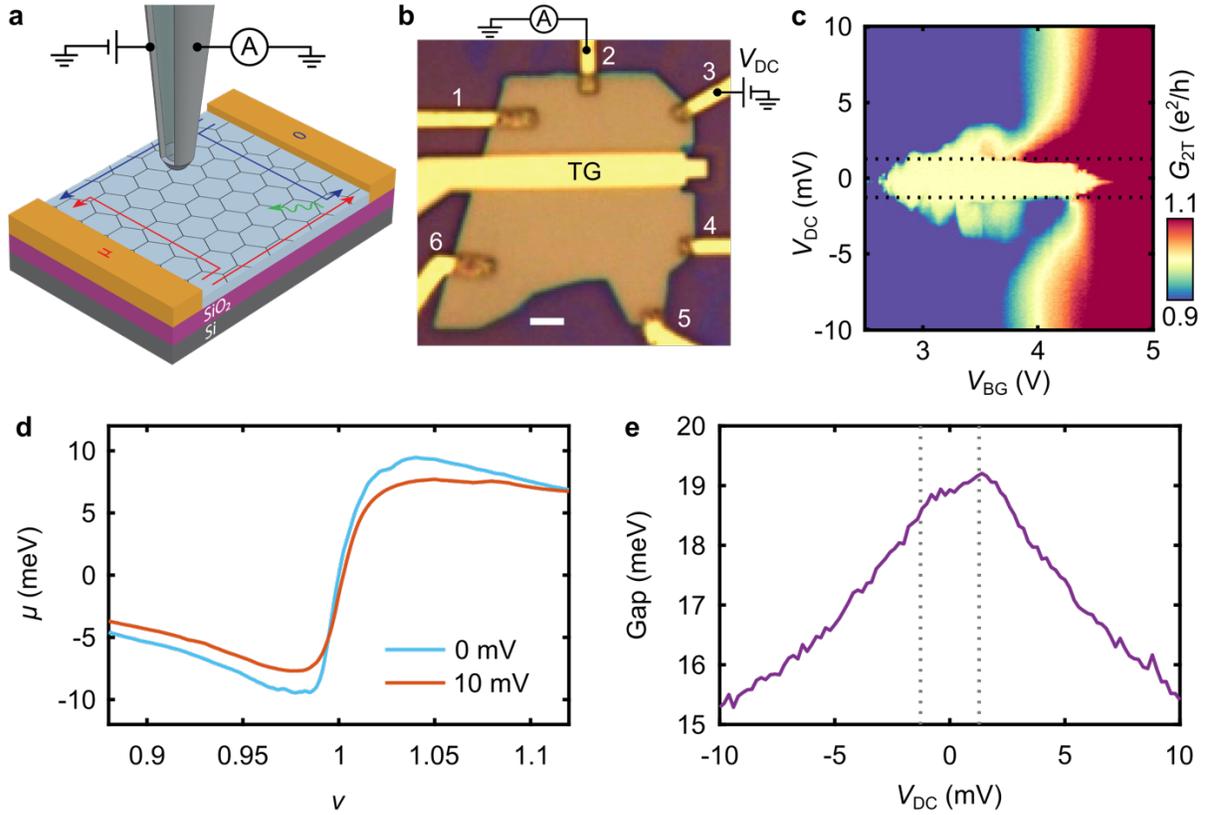

**Fig. 1 | Device characterization and ν=1 sensitivity to magnons. a,** Schematic of the experimental setup. Red and blue arrows denote the hot and cold quantum Hall edge states, respectively. Green curve denotes magnon generation for $\mu < -E_Z$. **b,** Optical micrograph of the hBN-encapsulated monolayer graphene device (scale bar 2 μm). TG denotes top gate. **c,** Two-terminal conductance $G_{2T}$ near the ν=1 plateau measured at 11 T between contacts 2 and 3 with the zero volts applied to the top gate. The plateau breaks down principally around $\pm E_Z$. **d,** $\mu(\nu)$ measured at 11 T in the bulk near contact 5 at $V_{DC}$=0 mV and 10 mV. The gap, taken as the peak excursion, is suppressed in the case of $V_{DC}$=10 mV. **e,** Bias-dependent energy gap extracted from chemical potential measurements as in **d**. The gap begins to reduce near $\pm E_Z$ marked by the gray dotted lines.

# Figure 2

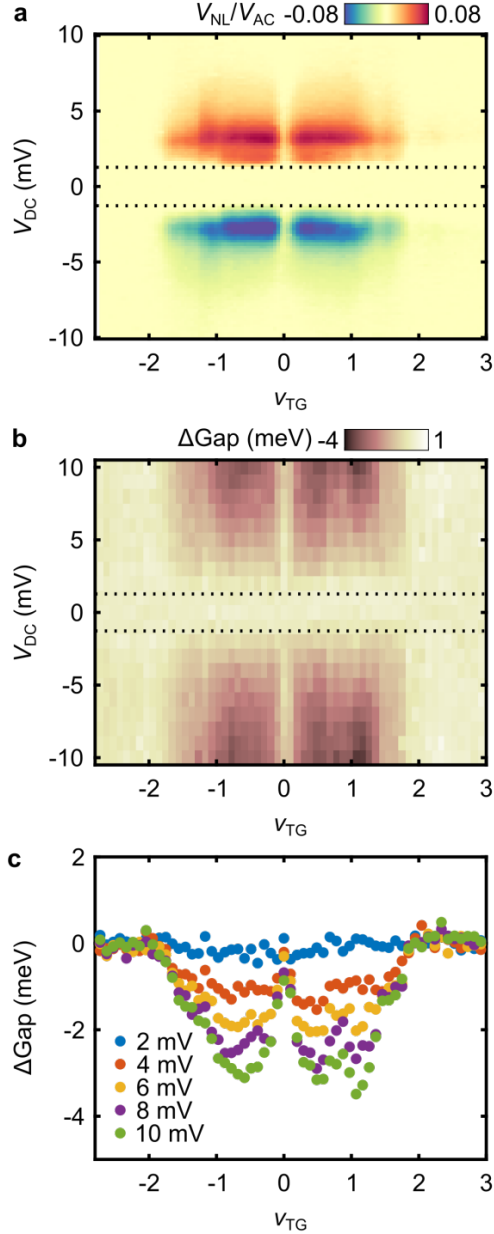

**Fig. 2 | Nonlocal magnon transport and gap suppression. a,** AC nonlocal voltage $V_{NL}$ measured between contacts 5 and 6 as a function of top gate filling factor $\nu_{TG}$ and DC bias $V_{DC}$ applied across contacts 2 and 3. A nonlocal voltage appears near $\pm E_Z$ (black dotted lines) in accordance with the standard picture of magnon transport. **b,** Change in the measured $\nu=1$ energy gap as a function of $V_{DC}$ and $\nu_{TG}$. ΔGap is calculated at each point by subtracting the average of the gap values at $|V_{DC}|<E_Z$. As in the case of $V_{NL}$, changes are only observed for $E_Z<|V_{DC}|$ and $0<|\nu_{TG}|<2$. **c,** Line traces from **b** showing the sharp disappearance of gap suppression near $\nu_{TG}=0$.

# Figure 3

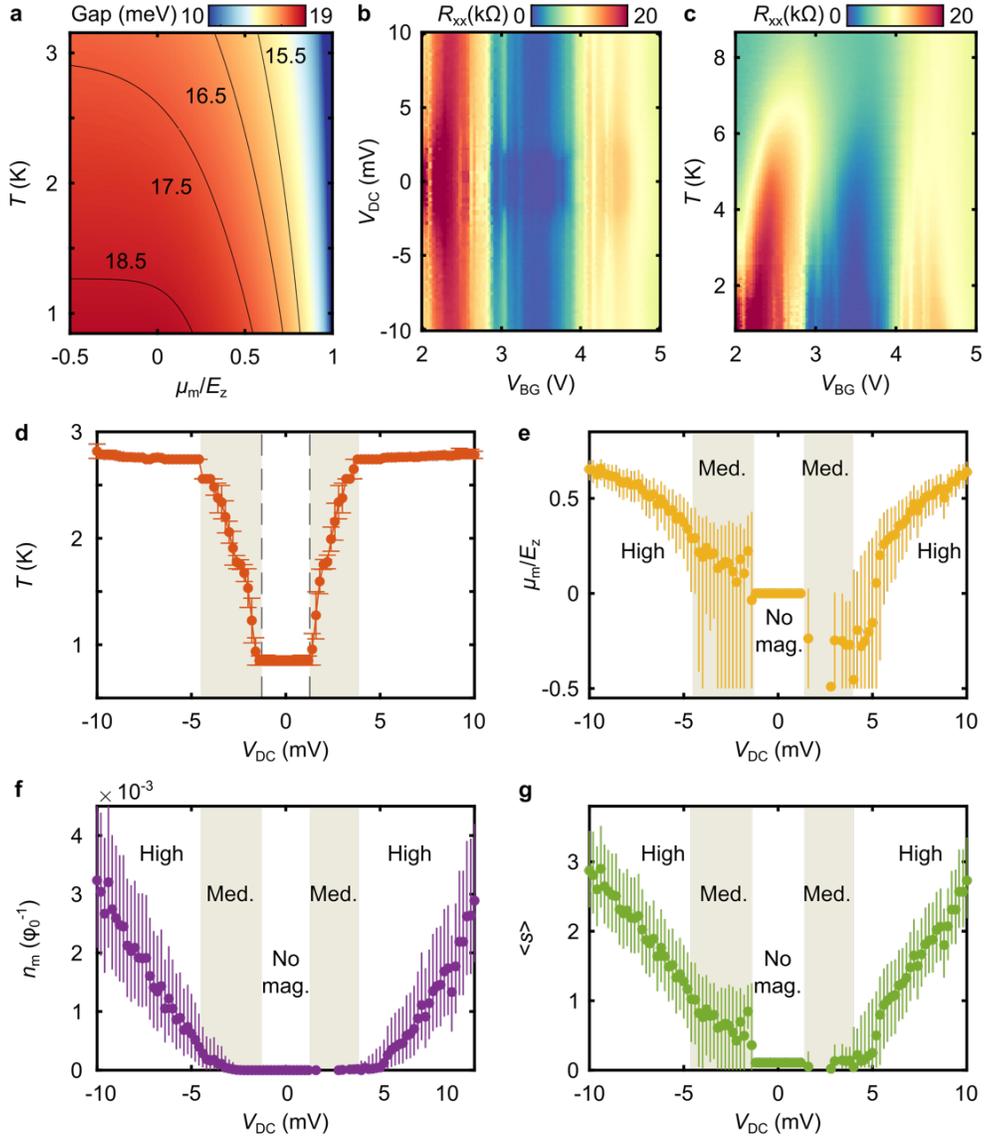

**Fig. 3 | Thermodynamics of free and bound magnons. a**, ν=1 gap as a function of magnon chemical potential $\mu_m/E_z$ and temperature $T$ computed using the skyrmion model. **b**, $R_{xx}$ as a function of $V_{DC}$ applied to contact 3 near ν=1 (see Extended Data Fig. 1a for circuit). The center of the ν=1 plateau is around $V_{BG}$=3.5 V. **c**, $R_{xx}$ as a function of temperature with no bias applied to contact 3 near ν=1 using the same circuit as **b**. **d**, Temperature of the system as a function of $V_{DC}$ extracted from $R_{xx}$ thermometry measurements (see Methods). Grey dashed lines mark the Zeeman energy. **e-g**, Magnon chemical potential $\mu_m/E_z$ (**e**), free magnon density per flux $n_m$ (**f**) and the number of extra flipped spin carried by the charge <$s$> (**g**) extracted from the skyrmion model (see Methods). The shaded region corresponds to a medium-bias regime where heating due to the magnon injection plays a key role.

# Figure 4

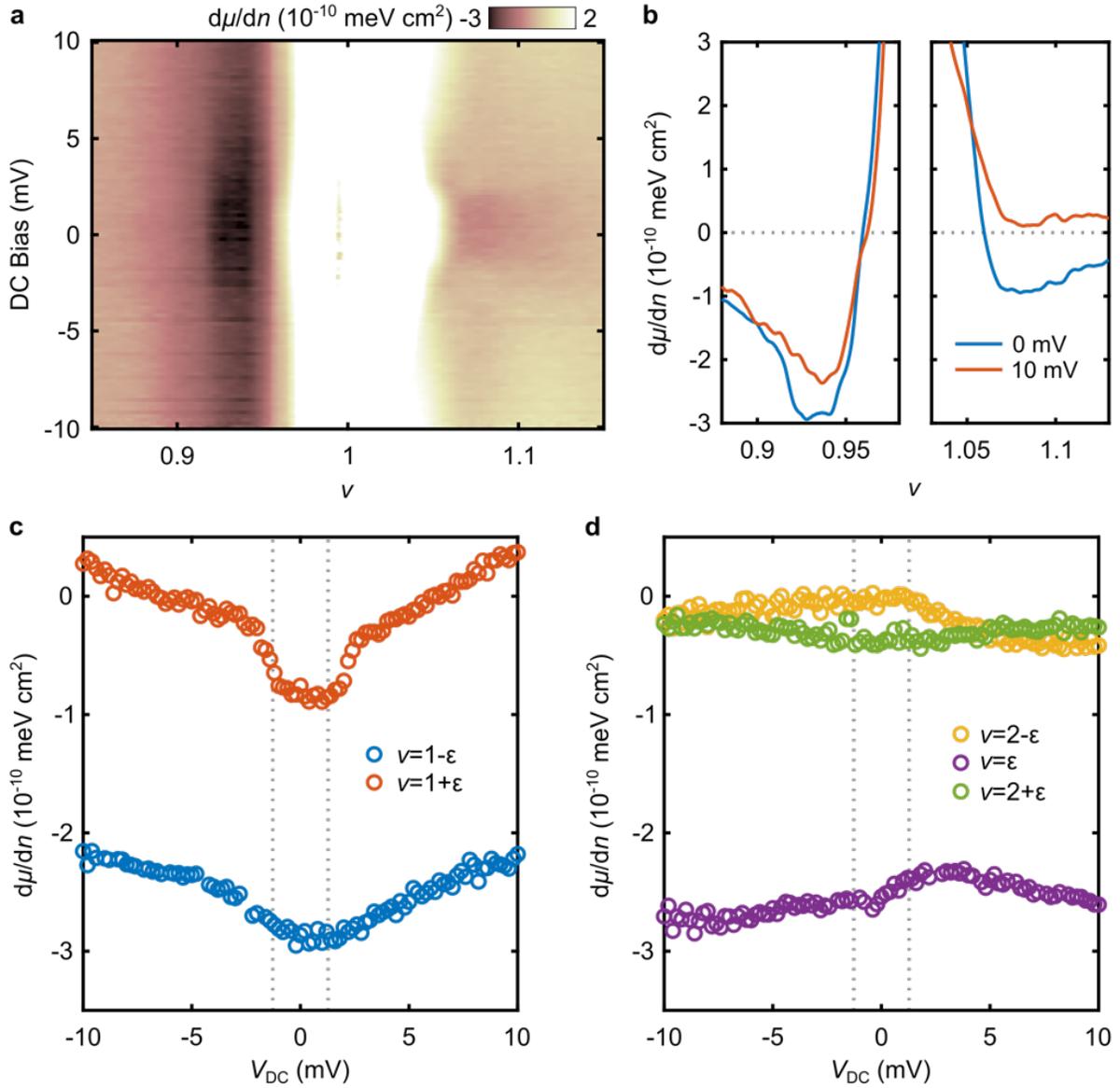

**Fig. 4 | Suppression of negative compressibility by the presence of magnons. a**, $d\mu/dn$ near ν=1 measured as a function of $V_{DC}$. **b**, Representative $d\mu/dn$ traces on the hole (left) and electron (right) sides of ν=1 measured with $V_{DC}$ =0 mV (blue) and $V_{DC}$ =10 mV (red) showing that the negative compressible states are suppressed by the presence of magnons. **c**, Averaged $d\mu/dn$ on the hole (blue) and electron (red) sides of ν=1 as a function of $V_{DC}$. The grey dotted lines mark the Zeeman energy $\pm E_z$. **d**, Average $d\mu/dn$ on the other WC states as a function of DC bias, showing no suppression by the $V_{DC}$. The grey dotted lines mark the Zeeman energy $\pm E_z$.

# Extended Data Figure 1

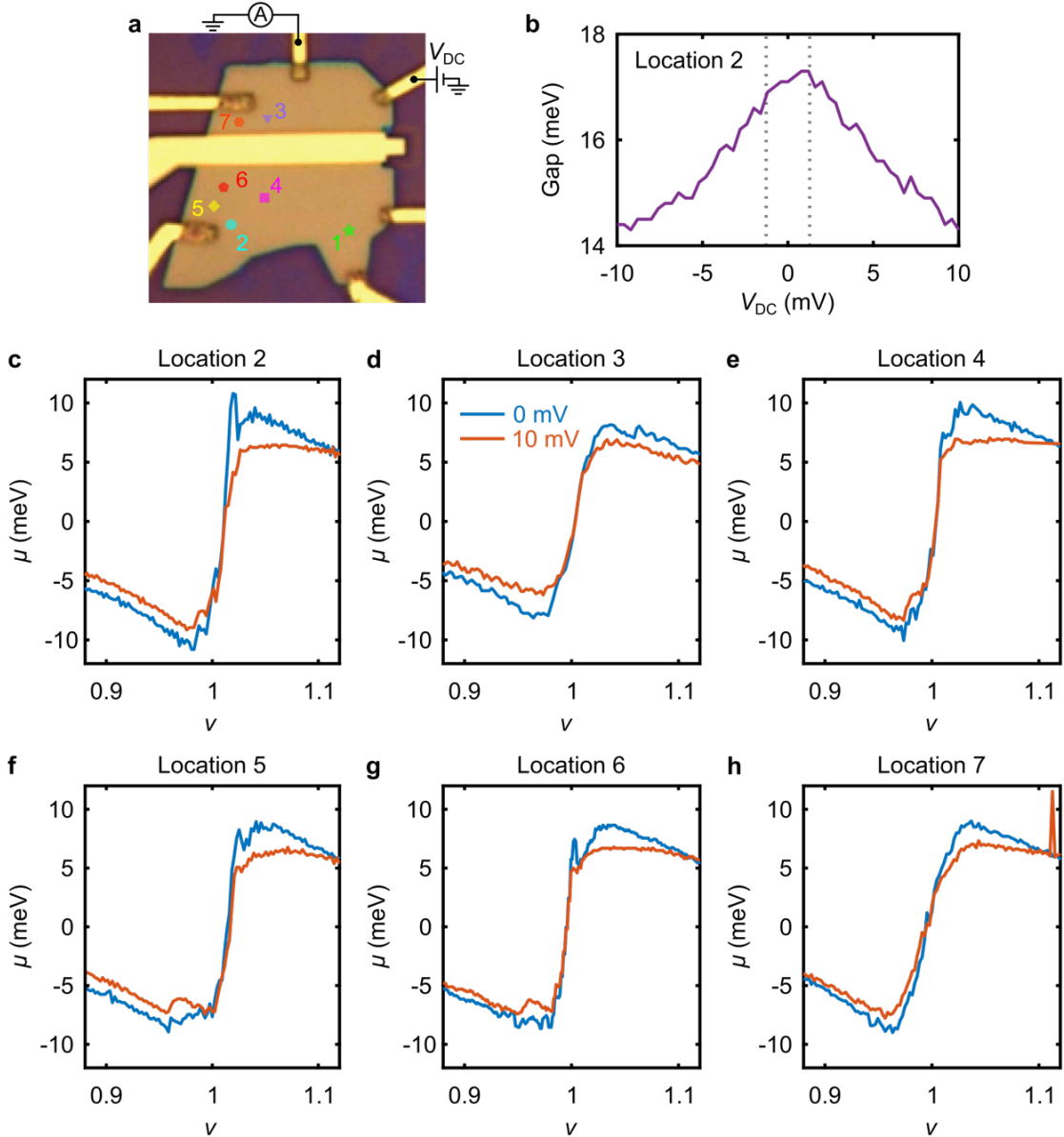

**Extended Data Fig. 1 | Additional examples of the ν=1 gap suppression by the presence of magnons. a,** Optical image of the device indicating the circuit used for magnon generation and the locations where the gap measurements were taken. **b,** Bias-dependent energy gap measured at location 2. The grey dotted lines mark $\pm E_Z$. While the origin of the small asymmetry for $|V_{DC}|<E_Z$ is unclear, its magnitude is much smaller than overall suppression observed at higher bias, and the top gate dependence shows that the onset consistently occurs near $E_Z$ (see Fig. 2). **c-g,** Chemical potential $\mu$ near ν=1 measured with $V_{DC}$=0 mV (blue) and $V_{DC}$=10 mV (red) at 6 different locations. Although the local value of the ν=1 gap varies, its reduction by the presence of magnons is clearly reproduced in all the data sets.

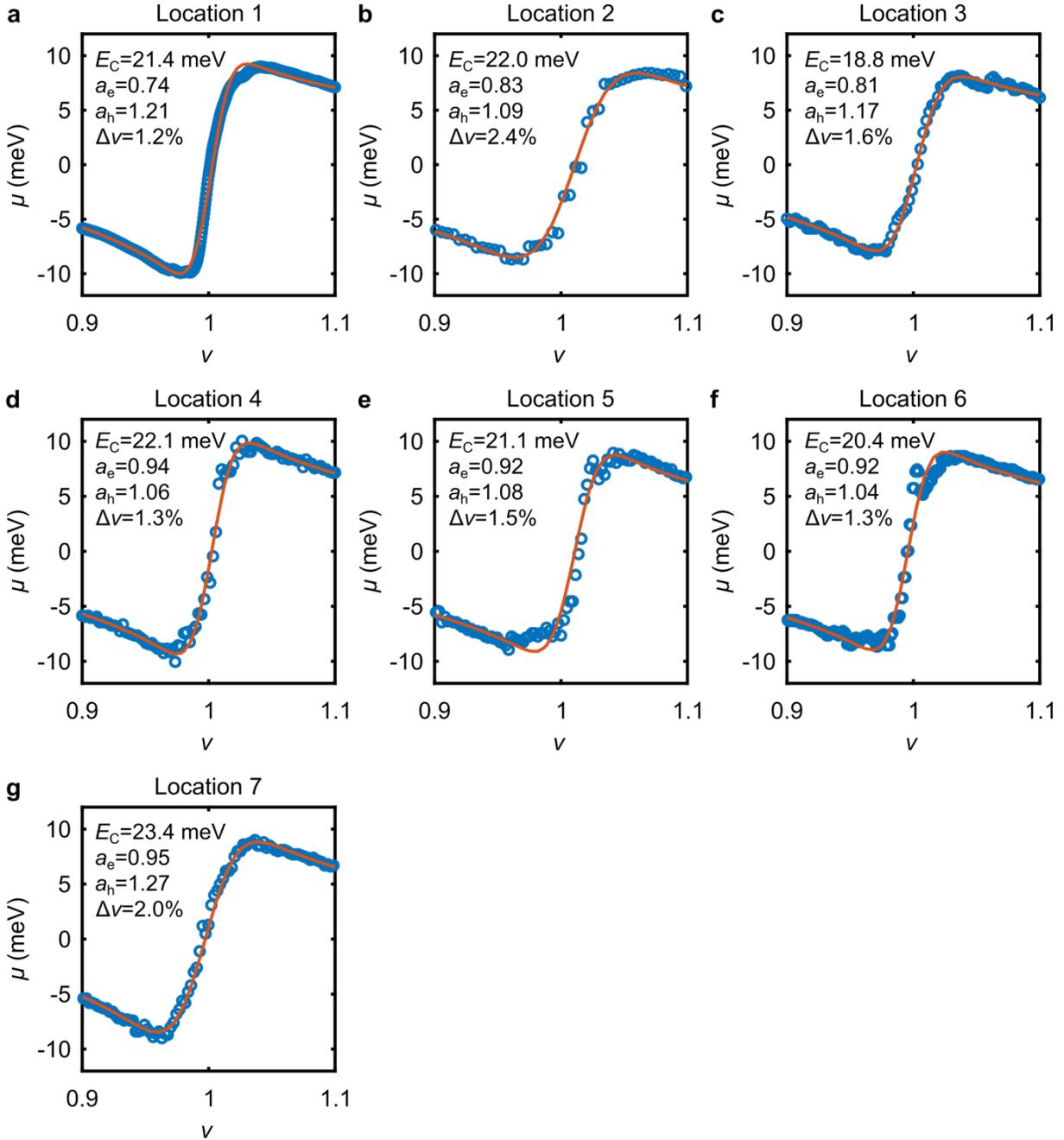

**Extended Data Fig.2 | Fitting of the chemical potential in the absence of magnons. a-g,** Chemical potential $\mu$ near $\nu=1$ measured with 0 mV (blue circles) DC bias applied to contact 3 and the best fit (red curves) using the skyrmion model by setting $\mu_m=0$ mV (see Methods) at 7 different locations. The values of $E_C$ obtained at these positions correspond to effective dielectric constants $\varepsilon$ ranging from 10.0 to 12.4.

# Extended Data Figure 3

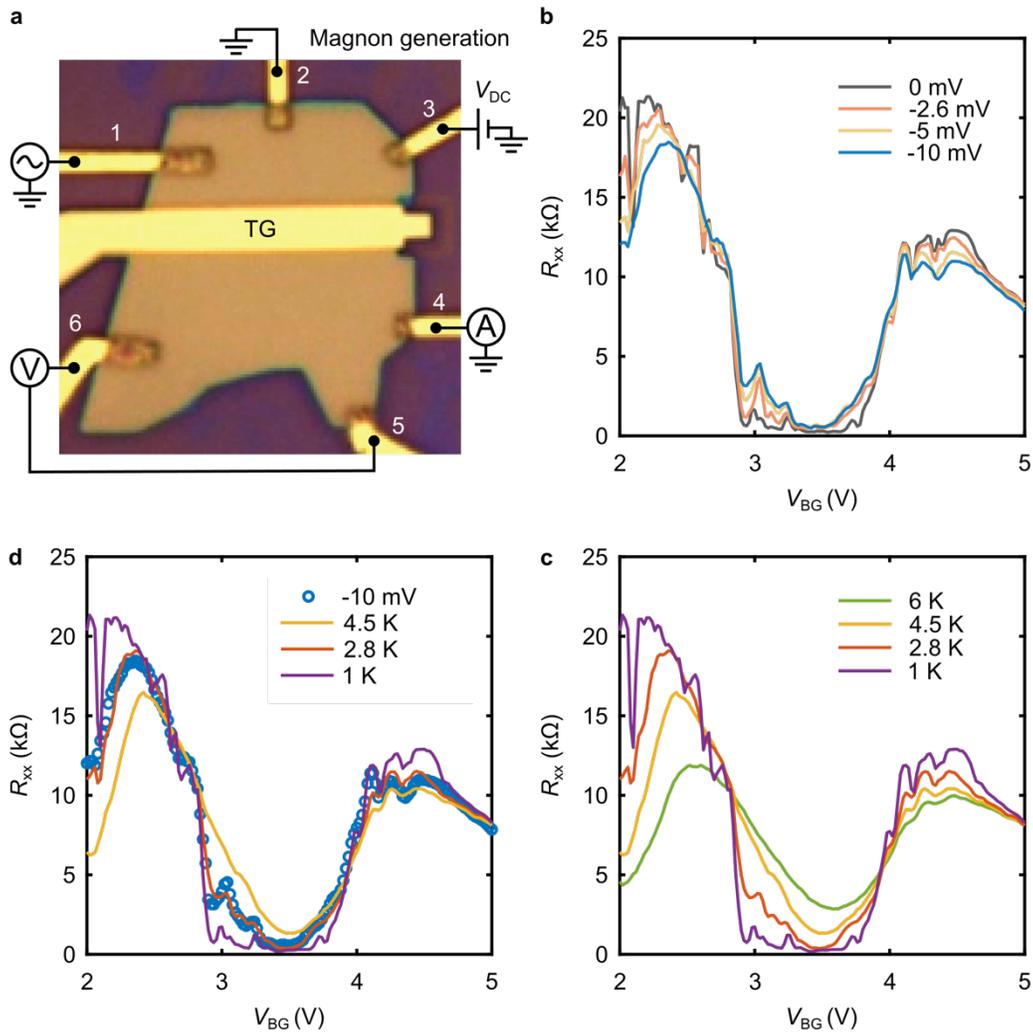

**Extended Data Fig. 3 | $R_{xx}$ thermometry. a,** Circuit used for $R_{xx}$ measurements. Contacts 2 and 3 are used to generate magnons. Contacts 1, 6, 5, and 4 are used to measure $R_{xx}$. The white arrows indicate the chirality of the current flow. **b,** Individual $R_{xx}$ traces measured at base temperature with various values of $V_{DC}$ applied to contact 3 near $\nu=1$ using the circuit shown in **a**. The center of the $\nu=1$ plateau is around a back gate voltage of 3.5 V. **c,** Individual $R_{xx}$ traces measured at various temperature with no bias applied to contact 3 near $\nu=1$ using the circuit shown in **a**. **d,** Individual $R_{xx}$ traces measured at base temperature with 10 mV applied to contact 3 (blue dots) and at various temperatures with 0 mV applied to contact 3 (orange, yellow and purple lines). The close agreement between the blue dotted line and the red line suggests that the effect of magnon generation on the $R_{xx}$ measurement is primarily due to heating. These measurements also demonstrate that the increase in temperature due to magnon generation does not exceed 4.5 K at $V_{DC}=-10$ mV.

# Extended Data Figure 4

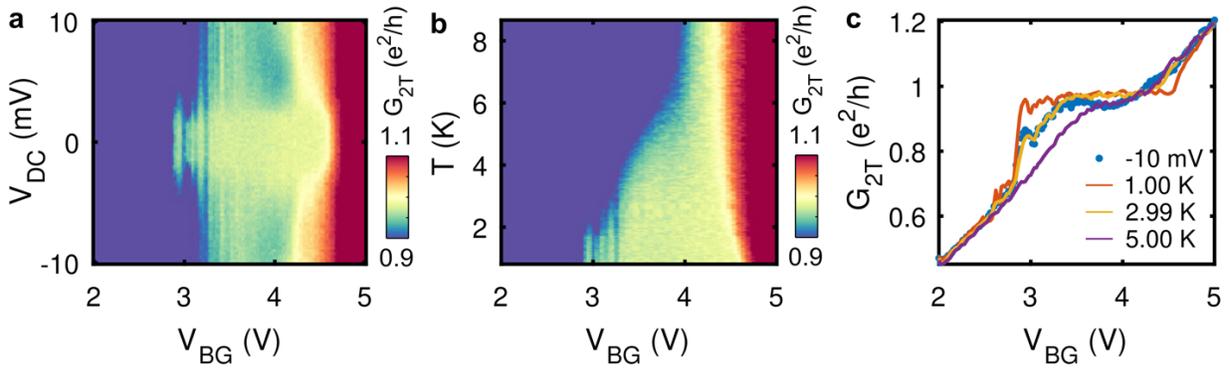

**Extended Data Fig. 4 | Alternative derivation of electron temperature using two-terminal conductance $G_{2T}$. a,** Bias-dependent two-terminal conductance $G_{2T}$ measured in the vicinity of the ν=1 plateau. **b,** Temperature-dependent $G_{2T}$ measured at zero bias in the same range of electron densities. **c,** $G_{2T}$ measured at -10 mV DC bias compared with selected zero-bias traces at elevated temperature.

# Extended Data Figure 5

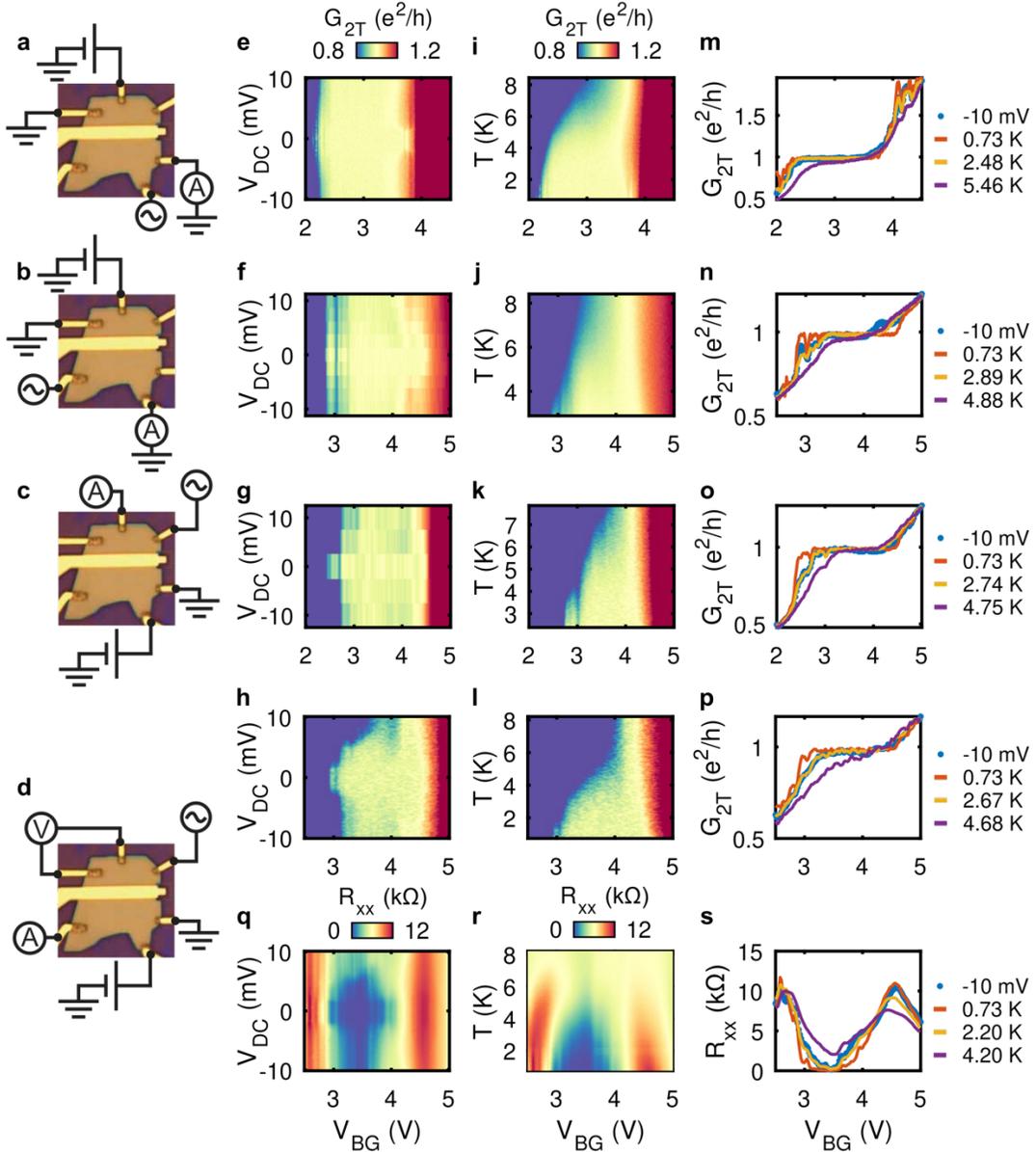

**Extended Data Fig. 5 | Temperature extraction from additional circuit configurations. a-d,** circuit configurations in which additional bias-dependent **(e-h)** and temperature dependent **(i-l)** two-terminal transport measurements were carried out. **m-p,** comparison of traces taken at $V_{DC}$=-10 mV and at base temperature with zero-bias traces taken at various temperatures. In each panel the middle value of temperature is that found to agree best with the $V_{DC}$=-10 mV trace in the least-squares sense. The good agreement between the -10 mV trace and the best-fit zero-bias trace indicates that the main effect of the bias in this circuit configuration is to elevate the temperature. **q-s,** additional $R_{xx}$ data acquired simultaneously with $G_{2T}$ using the circuit configuration shown in **d**. Estimation from $R_{xx}$ gives a slightly lower temperature than $G_{2T}$.

# Extended Data Figure 6

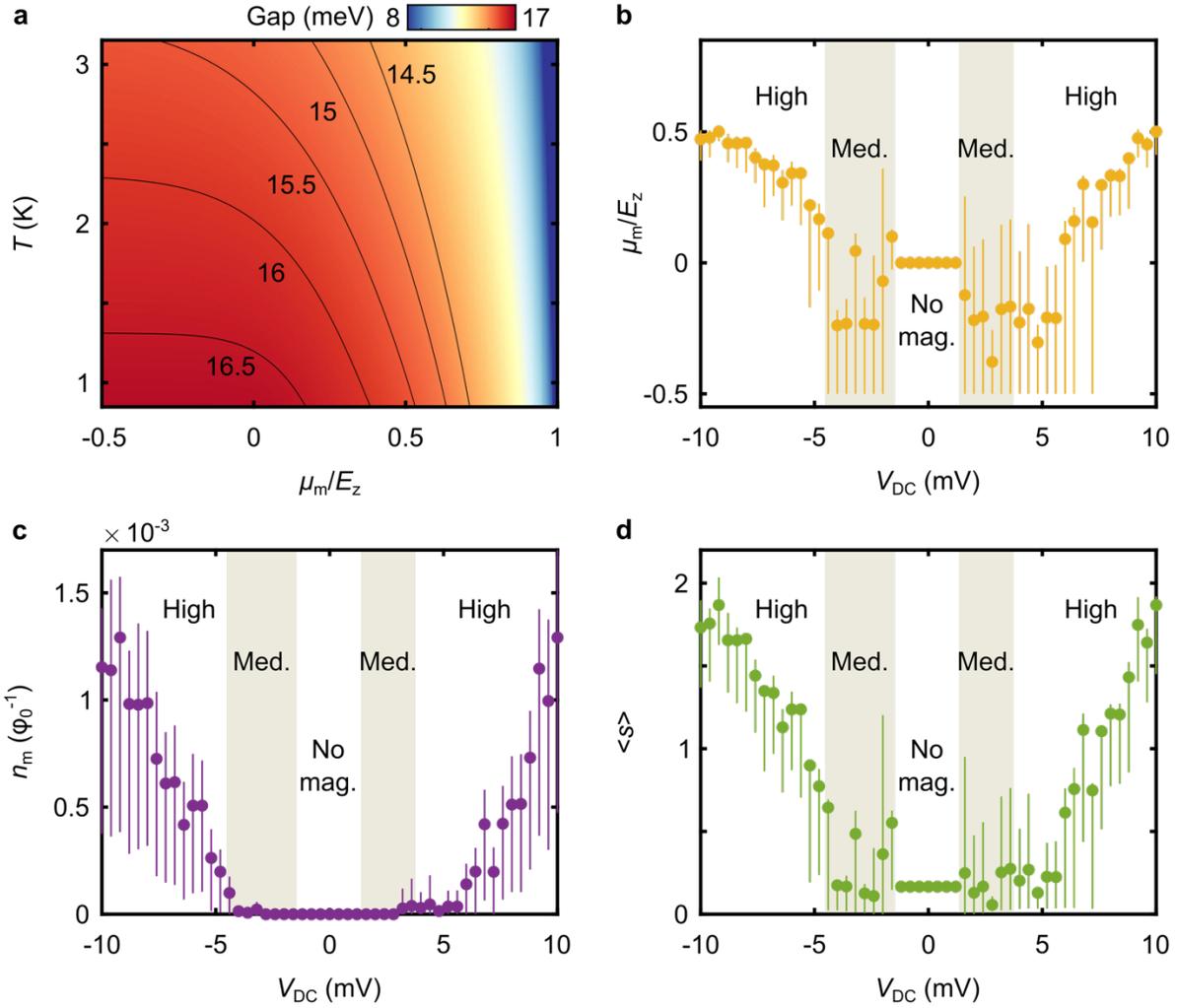

**Extended Data Fig. 6 | Thermodynamics of free and bound magnons extracted from location 2. a**, ν=1 gap as a function of magnon chemical potential $\mu_m/E_z$ and temperature $T$ computed using the skyrmion model. **b-d,** Magnon chemical potential $\mu_m/E_z$ (**b**), free magnon density per flux $n_m$ (**c**) and the mean number <s> of extra flipped spins carried by a charge (**d**), extracted from the skyrmion model (see Methods). The shaded region corresponds to a medium bias regime where heating due to magnon injected plays a key role.

# Extended Data Figure 7

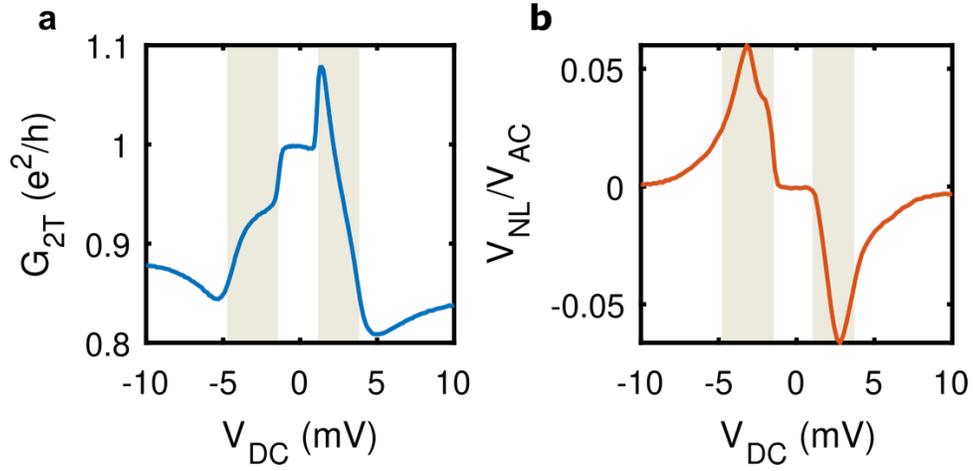

**Extended Data Fig. 7 | Three regimes in magnon transport characteristics. a,** $G_{2T}$ averaged over values of $V_{BG}$ on the ν=1 plateau as a function of DC bias. **b,** $V_{NL}/V_{AC}$ averaged over values of $ν_{TG}$ for $0<ν_{TG}<2$. On each plot, the low-, medium- and high-bias regimes are indicated by shading in the same manner as Fig. 3.

# Extended Data Figure 8

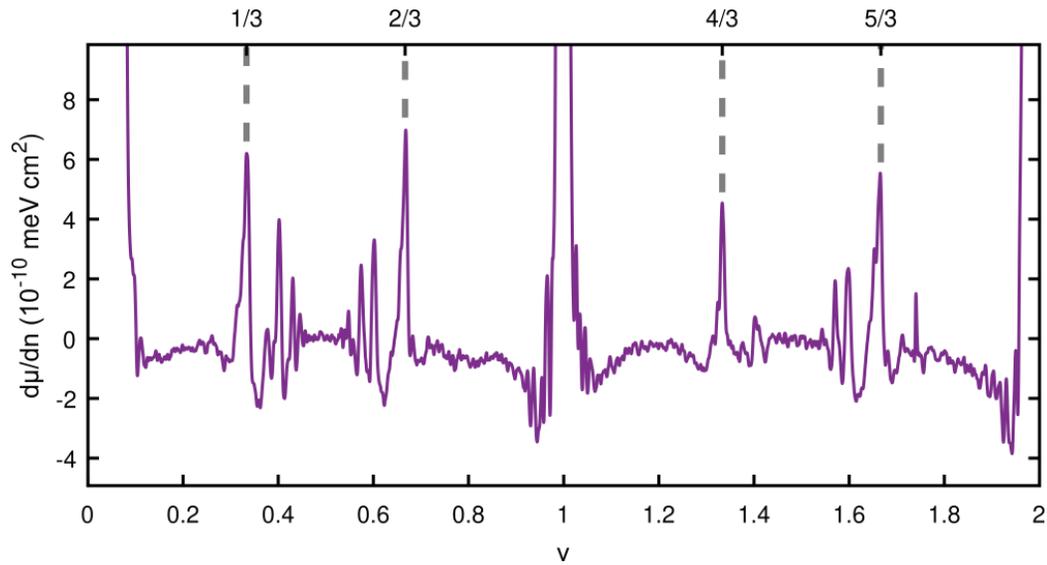

**Extended Data Fig. 8 | Robust ν=5/3 state.** Local inverse compressibility $d\mu/dn$ measured for 0< ν <2. In contrast to local compressibility studies performed on suspended monolayer graphene devices, a prominent peak at ν=5/3—comparable in strength to those at 1/3 and 2/3, and stronger than that at 4/3—is evident, suggesting that valley skyrmion formation in the device is *disfavored*.

# Extended Data Figure 9

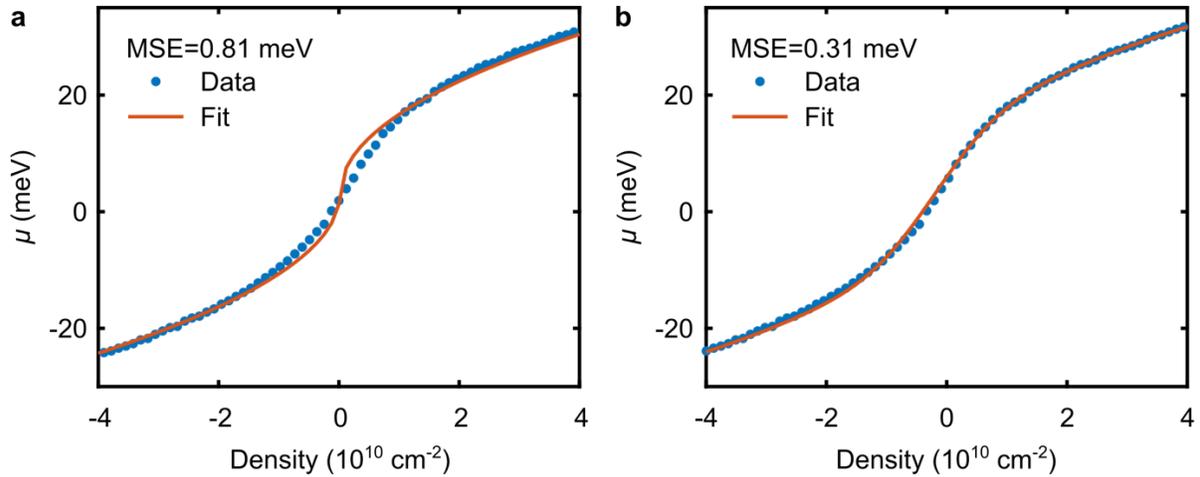

**Extended Data Fig. 9 | Zero-field fits to the Dirac point. a,** Fit comparing measured data to a model with no disorder broadening. The fit favors zero sublattice gap. **b,** Fit comparing measured data to a model with disorder broadening. The fit favors a scenario with a 12.3 meV sublattice gap with a disorder-broadening parameter of approximately $7\times10^9$ cm$^{-2}$, similar in magnitude to the broadening inferred from high-field compressibility measurements. The MSE of the broadened fit is improved compared to that of the unbroadened fit by more than a factor of two.